\documentstyle[11pt,newpasp,twoside,epsf]{article}
\def\edcomment#1{\iffalse\marginpar{\raggedright\sl#1\/}\else\relax\fi}
\reversemarginpar
\def\<<{{\ll}}
\def\>>{{\gg}}

\def\spose#1{\hbox to 0pt{#1\hss}}
\def\ltwig{\mathrel{\spose{\lower 3pt\hbox{$\mathchar"218$}}
     \raise 2.0pt\hbox{$\mathchar"13C$}}}
\def\gtwig{\mathrel{\spose{\lower 3pt\hbox{$\mathchar"218$}}
     \raise 2.0pt\hbox{$\mathchar"13E$}}}
\def\+/-{{\pm}}
\def\=={{\equiv}}
\def\Rstar{R_{\ast}}

\def\Mdot{\dot M}

\def\solar{\odot}
\def\Msun{M_{\solar}}

\newcommand{\beq}{\begin{equation}}
\newcommand{\eeq}{\end{equation}}

\begin{document}
\title{The Effects of Magnetic Fields on Line-Driven Hot-Star Winds}
 \author{Asif ud-Doula}
\affil{Department of Physics, North Carolina State University,
Raleigh, NC 27695-8202, USA}
\author{Stanley Owocki}
\affil{Bartol Research Institute, University of Delaware, Newark,
DE 19716, USA}

\begin{abstract}
This talk summarizes results from recent MHD simulations of
the role of a dipole magnetic field in inducing large-scale
structure in the  line-driven stellar winds of hot, luminous stars.
Unlike previous fixed-field
analyses, the MHD simulations here take full account of the
dynamical competition between the field and the flow. A key result
is that the overall degree to which the wind is influenced by the
field depends largely on a single, dimensionless `wind magnetic
confinement parameter', $\eta_\ast (= B_{eq}^2 R_{\ast}^2/\dot{M}
v_\infty$), which characterizes the ratio between magnetic field
energy density and kinetic energy density of the wind. For weak
confinement, $\eta_\ast \le 1$, the field is fully opened by wind
outflow, but nonetheless, for confinement as small as
$\eta_\ast=1/10$ it can have significant back-influence in
enhancing the density and reducing the flow speed near the
magnetic equator. For stronger confinement, $\eta_\ast > 1$, the
magnetic field remains closed over limited range of latitude and
height above the equatorial surface, but eventually is opened into
nearly radial configuration at large radii. Within the closed
loops, the flow is channeled toward loop tops into shock
collisions that are strong enough to produce hard X-rays. Within
the open field region, the equatorial channeling leads to oblique
shocks that are again strong enough to produce X-rays and also
lead to a thin, dense, slowly outflowing ``disk'' at the magnetic
equator.
\end{abstract}

\section{Introduction}

There is extensive evidence that hot-star winds are not
the steady, smooth outflows envisioned in the spherically
symmetric, time-independent models of  Castor, Abbott, and Klein (1975; hereafter CAK),
but instead have
extensive structure and variability on a range of spatial and
temporal scales. Relatively small-scale, stochastic structure --
e.g. as evidenced by often quite constant soft X-ray emission
(Long \& White 1980), or by UV lines with extended black troughs
understood to be a signature of a nonmonotonic velocity field
(Lucy 1982) -- seems most likely a natural result of the strong,
intrinsic instability of the line-driving mechanism itself (Owocki
1994; Feldmeier 1995). But larger-scale structure -- e.g. as
evidence by explicit UV line profile variability in even low
signal-to-noise IUE spectra (Kaper et al. 1996; Howarth \& Smith
1995) -- seems instead likely to be the consequence of wind
perturbation by processes occurring in the underlying star. For
example, the photospheric spectra of many hot stars show evidence
of radial and/or non-radial pulsation, and in a few cases there is
evidence linking this with observed variability in UV wind lines
(Telting, Aerts, \& Mathias 1997; Mathias et al. 2001).

An alternate scenario -- one explored  through dynamical
simulations in this talk -- is that, in at least some hot stars,
surface magnetic fields could perturb, and perhaps even channel,
the wind outflow, leading to rotational modulation of wind
structure that is diagnosed in UV line profiles, and perhaps even
to magnetically confined wind-shocks with velocities sufficient to
produce the relatively hard X-ray emission seen in some hot-stars.
The recent report by  Donati et al (2001) of a ca. 1000 G
dipole field in $\theta^1$~Ori~C suggests that, despite the lack of strong
convection
zones, hot stars can indeed have  magnetic fields.

\par
The focus of the present paper is to carry out MHD simulations of
how such magnetic fields on the surface of hot stars can influence
their radiatively-driven wind. Our approach here
represents a natural extension of the previous studies by Babel \&
Montmerle (1997a,b; hereafter BM97a,b), which effectively {\it
prescribed} a fixed magnetic field geometry to channel the wind
outflow. For large magnetic loops, wind material from opposite
footpoints is accelerated to a substantial fraction of the wind
terminal speed (i.e. $\sim 1000$~km/s) before the channeling
toward the loop tops forces a collision with very strong shocks,
thereby heating the gas to temperatures ($10^7-10^8$~K) that are
high enough to emit hard (few keV) X-rays. This `magnetically
confined wind shock' (MCWS) model was initially used to explain
X-ray emission from the Ap-Bp star IQ Aur (BM97a), which has a
quite strong magnetic field ($\sim 4$kG) and a rather weak wind
(mass loss rate $\sim 10^{-10} M_{\odot}/$~yr), and thus can
indeed be reasonably modeled within the framework of prescribed
magnetic field geometry.

Later, BM97b applied this
model to explain the periodic variation of X-ray emission of the
O7 star $\theta^1$~Ori~C, which has a much lower magnetic field
($\ltwig 1000$ G) and significantly stronger wind (mass loss rate
$\sim 10^{-7} M_{\odot}/$~yr), raising now the possibility that the
wind itself could influence the field geometry in a way that is
not considered in the simpler fixed-field approach.
\par
The simulation models here are based on an isothermal
approximation of the complex energy balance, and so can provide
only a rough estimate of the level of shock heating and X-ray
generation. But a key advantage over previous approaches is that
these models do allow for such a fully dynamical competition
between the field and flow. A central result is that the overall
effectiveness of magnetic field in channeling the wind outflow can
be well characterized in terms of single `wind magnetic
confinement parameter' $\eta_\ast$, defined below (\S 2).
In \S 3 we discuss the key results of our simulations,
and in \S 4 we summarize our main conclusions.

\section{The Wind Magnetic Confinement Parameter}
As detailed in ud-Doula and Owocki (2002),
the relative effectiveness of magnetic fields in confining
and/or channeling a wind outflow depends largely on the ratio
of energy densities in the field vs. flow
\begin{eqnarray}
\eta (r, \theta) &\equiv& \frac{B^2/8\pi}{\rho v^2/2}
\approx \frac{B^2 r^2}{\dot{M}v} \label{etadef}
\\
&=& \left [\frac{B_{\ast}^2 (\theta )
{R_{\ast}}^2}{\dot{M}v_{\infty}}\right ] \left
[\frac{(r/R_{\ast})^{2-2q}}{1-R_{\ast}/r} \right ] \, . \nonumber
\end{eqnarray}
The latter approximations assume a separation in terms of the latitudinal
variation of the surface field, $B_\ast (\theta)$
(e.g. for dipole $B_\ast^2 (\theta) = B_o^2 ( \cos^2 \theta +
\sin^2 \theta/4 )$) and radial variation $B(r)
=B_{\ast}(R_{\ast}/r)^q$, with, e.g., for a simple dipole $q=3$.
The right square bracket in eqn (1) represents
the spatial variations of this energy ratio and the left square
bracket represents a dimensionless constant that characterizes the
overall relative strength of field vs. wind. Evaluating this at
 the magnetic equator ($\theta=90\deg$), where the
radial wind outflow is in  most direct
competition with  a horizontal orientation of the
field; we can thus define an equatorial `wind magnetic confinement
parameter',
\begin{eqnarray}
\eta_{\ast} &\equiv&
\frac{B_\ast^2 (90\deg) {R_{\ast}}^2} {\dot{M}v_{\infty}}
=0.4  \,  \frac{B_{100}^2 \, R_{12}^2}{\dot{M}_{-6} \, v_8}.
\label{wmcpdef}
\end{eqnarray}
where $\dot{M}_{-6} \equiv \dot{M}/(10^{-6}\, M_{\odot}$/yr),
$B_{100} \equiv B_o/(100$~G), $R_{12} \equiv
R_{\ast}/(10^{12}$~cm), and $v_{8} \equiv v_{\infty}/(10^8$~cm/s).
For a typical OB supergiant, e.g. $\zeta$ Pup, this suggests that
significant magnetic confinement or channeling of the wind should require
fields of order $ \sim 100$~G. By contrast, in the case of the
sun, with the much weaker mass loss (${\dot  M}_\odot \sim
10^{-14}~M_{\odot}$/yr)  global field
$B_{o} \sim 1$~G is sufficient to yield $\eta_{\ast} \simeq 40$,
implying a substantial magnetic confinement of the solar coronal
expansion. This is consistent with the observed large extent of
magnetic loops in optical, UV and X-ray images of the solar
corona.
\section{Results}
 We study the dynamical competition between field
and wind by evolving our MHD simulations from an initial condition
at time $t=0$, when a dipole  magnetic field is suddenly
introduced into a previously relaxed, 1D spherically symmetric CAK
wind for various assumed values of the wind magnetic
confinement parameter $\eta_{\ast}$. We first confirm that, for
sufficiently weak confinement, i.e., $\eta_{\ast} \le 0.01$,  the
wind is essentially unaffected by the magnetic field. But for
models within the range $1/10 < \eta_{\ast} < 10$, the field has a
significant influence on the wind. For our main parameter study,
the variations in $\eta_{\ast}$ are implemented solely through
variations in the assumed magnetic field strength, with the
stellar and wind parameters fixed at values appropriate to a
typical OB supergiant, e.g.  $\zeta$ Pup. Detailed analysis and results
of all these models are presented in ud-Doula and Owocki (2002), here we
merely summarize them.

\subsection{Global Wind Structure for Strong, Moderate, and Weak  Fields}

As an example, Figure 1 illustrates the global configurations of
magnetic field, density, and radial and latitudinal components of
velocity at the final time snapshot, $t=450$~ksec after initial
introduction of the dipole magnetic field. The top, middle, and
bottom rows show respectively results for a weak, moderate, and
strong field, characterized by confinement parameters of
$\eta_{\ast} =1/10$, $1$, and $10$.

For the weak magnetic case $\eta_{\ast}=1/10$, the flow
effectively extends the field to almost a purely radial
configuration everywhere. Nonetheless, the field
still has a noticeable influence, deflecting the flow slightly
toward the magnetic equator (with peak latitudinal speed
$\max(v_\theta) \simeq 70$~km/s) and thereby leading to an
increased density and a decreased radial flow speed in the
equatorial region. With the increase of $\eta_{\ast}$,
this equatorward deflection becomes more pronounced, with a faster
latitudinal velocity component ($\max(v_\theta ) \simeq
300$~km/s for $\eta_{\ast}=1$, and $> 500$~km/s
for $\eta_{\ast}10$), and a correspondingly stronger equatorial change in
density and radial flow speed leading to a quite narrow equatorial ``disk'' of
dense, slow outflow. For the strong magnetic case
$\eta_{\ast}=10$, the near-surface
fields now have a closed-loop configuration out to a substantial
fraction of a  stellar radius above the surface, but outside and well
above the  closed region, the flow is quasi-steady.

\subsection{Variability of Near-Surface Equatorial Flow}

In contrast to this relatively steady, smooth nature of the outer
wind, the flow near the star can be quite structured and variable
in the equatorial regions.
Figure 2 zooms in on the near-star equatorial region,
comparing density (upper row, contours), mass flux (arrows), and
field lines (lower row) at an arbitrary time snapshot long after
the initial condition ($t >400$~ksec), for three models with magnetic
confinement numbers $\eta_{\ast}=$ 1, $\sqrt{10}$, and 10.

For all three cases, high-density knots appear at few tenths of $\Rstar$
above the equatorial surface. These knots are too dense for the
the radiative-driving to maintain a net outward acceleration against the
inward pull of the stellar gravity. As such, they eventually
fall back onto the stellar surface. Although strong magnetic tension
can support these knots for a while, it eventually forces them to
slide to north or south hemispheres.
The animations of the time evolution for these models
show that such infalls occur at semi-regular intervals of
about 200~ksec.

\section{Conclusion}
Observational implications of these MHD simulations
are discussed in ud-Doula and Owocki (2002).
Here, we merely highlight some of our conclusions.

We found that
the general effect of magnetic field in channeling the
stellar wind depends on the overall ratio of magnetic to
flow-kinetic-energy density, as characterized by the wind magnetic
confinement parameter, $\eta_{\ast}$, defined here in eqn.
(2).
For moderately small confinement, $\eta_{\ast}=1/10$, the
wind extends the surface magnetic field into an open, nearly
radial configuration. But even at this level, the field still has
a noticeable global influence on the wind, enhancing the density
and decreasing the flow speed near the magnetic equator.
For intermediate confinement, $\eta_{\ast}=1$, the fields
are still opened by the wind outflow, but near the surface retain
a significant non-radial tilt, channeling the flow toward the
magnetic equator with a latitudinal velocity component as high as
300~km/s.
On the other hand, for strong confinement, $\eta_{\ast}=10$, the field remains
closed in loops near the equatorial surface. Wind outflows
accelerated upward from opposite polarity footpoints are channeled
near the loop tops into strong collision, with characteristic
shock velocity jumps of up to about 1000~km/s, strong enough to
lead to hard ($>~1$~keV) X-ray emission.

In contrast to  previous steady-state, fixed-field
models (e.g. BM97a), the time-dependent dynamical models here indicate that
stellar gravity pulls the compressed, stagnated material within
closed loops into an infall back onto the stellar surface, often
through quite complex, intrinsically variable flows that follow
magnetic channels randomly toward either the north or south loop
footpoint.

\def\blankline{\par\vskip \baselineskip}
\blankline \noindent{\it Acknowledgements.} This research was
supported in part by NASA grant NAG5-3530 and NSF grant
AST-0097983 to the Bartol Research Institute at the University of
Delaware. A. ud-Doula acknowledges support of NASA's Space Grant
College program at the University of Delaware.

\section*{Discussion}
\noindent {\it Brown:} Am I right in thinking you have no rotation
in the results you showed, which could prevent fall back? Also,
your results are $\Mdot=10^{-6} \Msun/yr$ which is much larger
than for Be star winds ($\sim 10^{-9}\Msun/yr$),
so, it is not surprising that even very large fields have limited effect.\\

\noindent {\it ud-Doula:} Yes, you are right. I did not include
rotation in the models I presented here. But I did do a parameter
study that included rotation, and those models will be presented
by Stan Owocki later in the day.

One important point I wanted to make in this talk is that the mass
loss rate alone doesn't determine if magnetic fields can play an
important role or not, it is the combination of parameters that
define $\eta_\ast$ that determines the importance of the field. In
general, low mass loss rate implies high value of $\eta_\ast$ even
for moderate field strengths (10-100 G), and this is particularly
true for Be stars.
So modest magnetic fields could play a significant role in Be stars.
\\
\noindent {\it Wade:} You made the point that your models predict
a disc around $\theta^1 Ori C$ that is highly variable. However,
the optical spectroscopic diagnostics that probe the disk (in
particular, the $H\alpha$ and $H\beta$ and He 4686 emission lines)
show that it is stable over many, many rotations. This seems to be
telling us that the disk is quite stable. Can you reconcile these
results?
\\
\noindent {\it ud-Doula:} The variability I am suggesting is mostly
confined to region near the surface
of the star, the flow far away from the surface, including the disk,
may be quasi-steady. \\
\noindent {\it Steinitz:} Your MHD equations did not include an
essential mechanism: the DME (Diamagnetic Effect). Inclusion may
change your fitting parameters drastically since it is not clear
that exclusion of DME describes anything close to real physics.
\\
\noindent {\it ud-Doula:} Indeed I have not included DME
explicitly in my MHD equations. As such, I do not know how much
DME can influence my results, but this is something
I should, perhaps, look into in the near future.\\
\noindent {\it Ignace:} I know that your models do not include
shock generation in the wind at large distances, but do you think
that the hot gas in the equatorial region from the
magnetically-guided self-colliding wind will dominate the X-ray
emission?
\\
\noindent {\it ud-Doula:} I think you are right, it will dominate the hard X-ray emission.
As for the soft X-ray, likely the result of the strong instabilities
in the wind, can be formed at large distances.

\begin{figure}
\begin{center}
\plotone{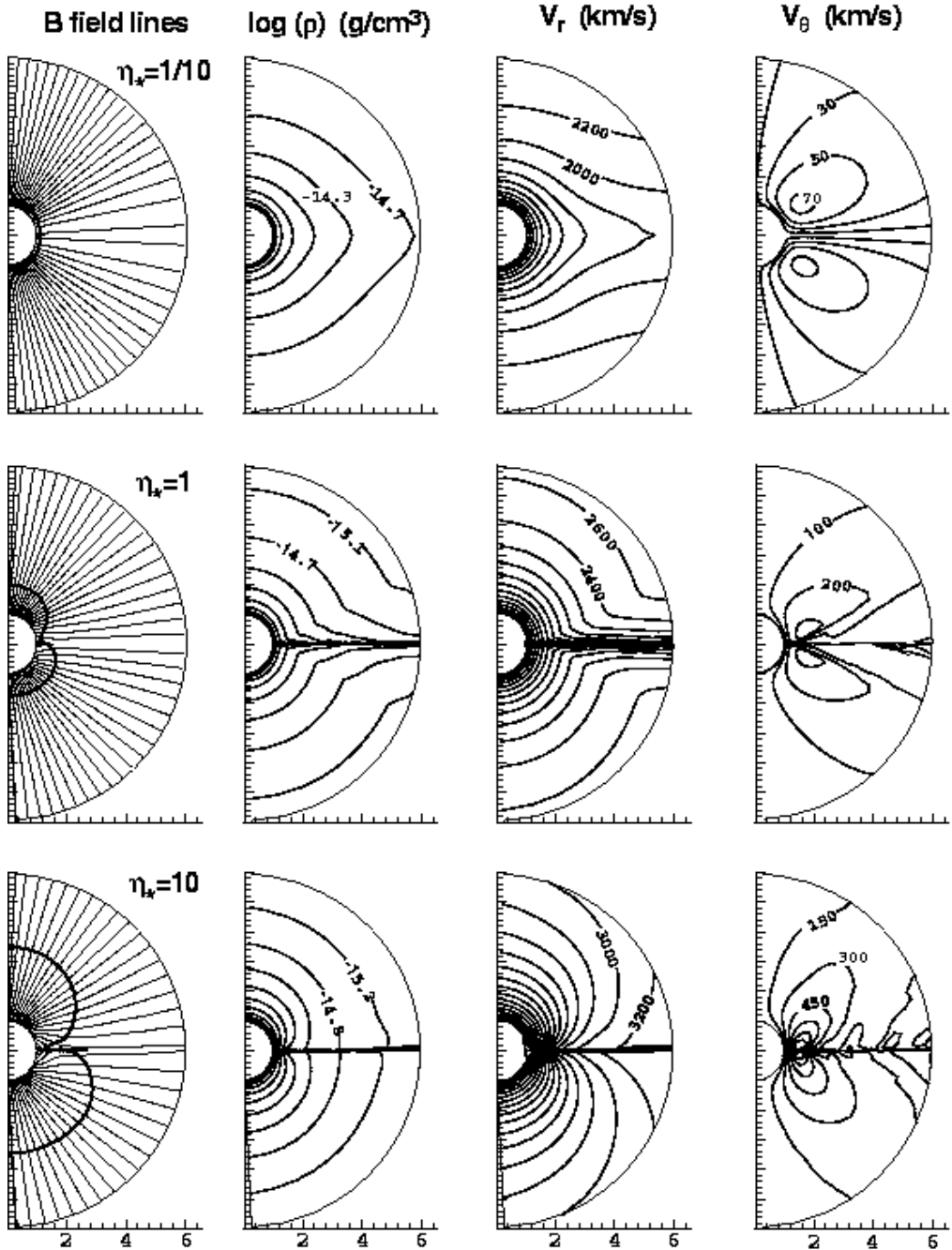}
\end{center}
\caption
{
Comparison of overall properties at the final simulation
time ($t=450$~sec) for 3 MHD models, chosen to span a range of
magnetic confinement from
small (top row; $\eta_{\ast}=1/10$),
to medium (middle row; $\eta_{\ast}=1$),
to large (bottom row; $\eta_{\ast}=10$).
The leftmost panels show magnetic field lines, together with the
location (bold contour) of the Alfven radius, where the radial flow speed
equals the Alfven speed.
From left to write, the remaining columns show
contours of log(density), radial velocity, and latitudinal velocity.
}
\label{fig3}
\end{figure}
\begin{figure}
\begin{center}
\plotone{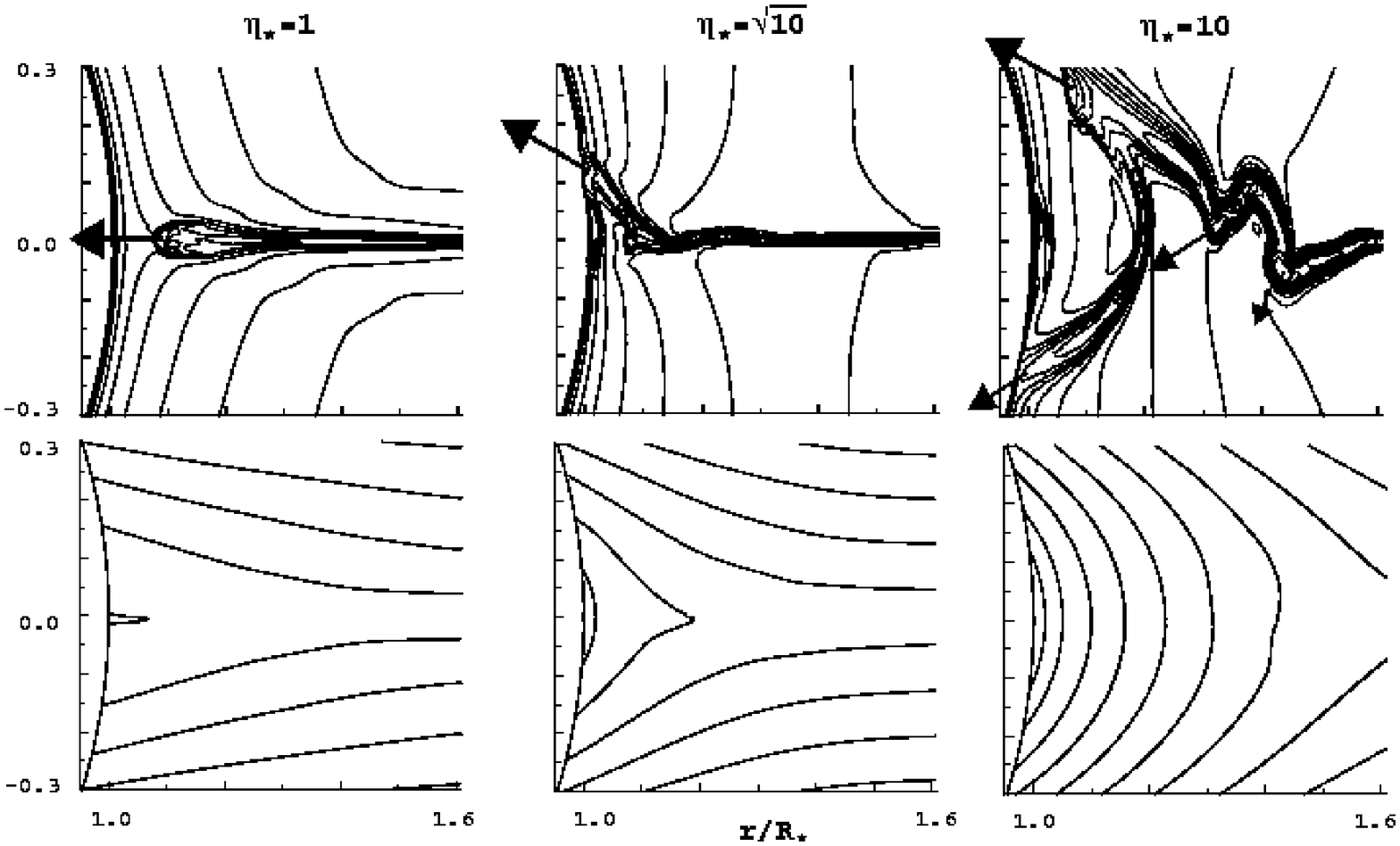}
\end{center}
\caption
{
Contours of log(density) (upper row) and magnetic field lines (lower
row) for the inner, magnetic-equator regions of
MHD models with moderate ($\eta_{\ast}=1$; left),
strong ($\eta_{\ast}=\sqrt{10}$; middle), and
strongest ($\eta_{\ast}=10$; left) magnetic confinement,
shown at a fixed,  arbitary time snapshot well after ($t \ge 400$~ksec)
the initial condition.
The arrows represent the direction and magnitude of the mass flux, and
show clearly that the densest structures are undergoing an infall back
onto the stellar surface.
}
\label{fig4}
\end{figure}


\begin{references}
Babel, J.~\& Montmerle, T.\
1997, \aap, 323, 121

Babel, J. \& Montmerle, T.
1997, \apjl, 485,L29

Castor, J.~I., Abbott, D.~C., \& Klein, R.~I.\ 1975, \apj, 195, 157\

Donati, J.-F., Wade, G.~A.,
Babel, J., Henrichs, H.~F., de~Jong, J.~A., Harries, T.~J. 2001,
MNRAS, 326, 1265

Feldmeier, A.\ 1995, \aap, 299, 523

Howarth,
I.~D.~\& Smith, K.~C.\ 1995, \apj, 439, 431

Kaper, L., Henrichs,
H.~F., Nichols, J.~S., Snoek, L.~C., Volten, H., \& Zwarthoed,
G.~A.~A.\ 1996, \aaps, 116, 257

Long, K.~S.~\& White, R.~L.\ 1980, \apjl, 239, L65

Lucy, L.~B. 1982, \apj, 255, 286

Mathias, P.,
Aerts, C., Briquet, M., De Cat, P., Cuypers, J., Van Winckel, H.,
Flanders., \& Le Contel, J.~M.\ 2001, \aap, 379, 905

Owocki, S.~P., Cranmer, S.~R., \& Blondin, J.~M.\ 1994, \apj, 424, 887

Telting, J.~H., Aerts, C., \& Mathias, P.\ 1997, \aap, 322, 493

ud-Doula, A. \& Owocki, S. 2002, \apj, 576, 413
\end{references}
\end{document}